\documentclass[conference]{IEEEtran}
\IEEEoverridecommandlockouts
\usepackage{cite}
\usepackage{amsmath,amssymb,amsfonts}
\usepackage{algorithmic}
\usepackage{graphicx}
\usepackage{textcomp}

\usepackage{url}
\usepackage{caption}
\usepackage{enumerate}
\usepackage{CJK}
\usepackage[normalem]{ulem}
\usepackage{enumitem} 
\usepackage{amsmath}

\usepackage{float}
\usepackage{amssymb}
\usepackage{pifont} 
\usepackage[utf8]{inputenc}
\usepackage{tabularx} 
\usepackage{multirow} 
\usepackage{graphicx} 
\usepackage{xcolor}
\usepackage{colortbl}
\usepackage{arydshln}
\usepackage{subcaption}
\usepackage{diagbox} 
\usepackage{booktabs}
\usepackage{listings}
\usepackage{threeparttable}

\def\BibTeX{{\rm B\kern-.05em{\sc i\kern-.025em b}\kern-.08em
    T\kern-.1667em\lower.7ex\hbox{E}\kern-.125emX}}

\begin{document}

\title{V-CASS: Vision-context-aware Expressive \\Speech Synthesis for Enhancing User Understanding of Videos\\
}

\author{
Qixin Wang\textsuperscript{1},
Songtao Zhou\textsuperscript{1},
Zeyu Jin\textsuperscript{1},
Chenglin Guo\textsuperscript{2},
Shikun Sun\textsuperscript{1},
Xiaoyu Qin\textsuperscript{1\thanks{Corresponding author: Xiaoyu Qin (xyqin@tsinghua.edu.cn)}}
\\[1ex]
\textsuperscript{1}\textit{Department of Computer Science and Technology, Tsinghua University, Beijing, China}\\
\textsuperscript{2}\textit{Stern School of Business, New York University, New York, USA}
\\[1ex]
\{wqx22, zst24, jinzeyu23\}@mails.tsinghua.edu.cn,\\
chenglin.guo@stern.nyu.edu, ssk21@mails.tsinghua.edu.cn, xyqin@tsinghua.edu.cn
}

\maketitle

\begin{abstract}
Automatic video commentary systems are widely used on multimedia social media platforms to extract factual information about video content. 
However, current systems may overlook essential para-linguistic cues, including emotion and attitude, which are critical for fully conveying the meaning of visual content. 
The absence of these cues can limit user understanding or, in some cases, distort the video's original intent. 
Expressive speech effectively conveys these cues and enhances the user's comprehension of videos. 
Building on these insights, 
this paper explores the usage of vision-context-aware expressive speech in enhancing users' understanding of videos in video commentary systems\footnote{The \textbf{demo} \& \textbf{appendix} of the paper are available at {\color{blue}\url{https://v-cass.github.io/V-CASS/}}.}. 
Firstly, our formatting study indicates that semantic-only speech can lead to ambiguity, and misaligned emotions between speech and visuals may distort content interpretation. 
To address this, we propose a method called vision-context-aware speech synthesis (V-CASS). 
It analyzes para-linguistic cues from visuals using a vision-language model and leverages a knowledge-infused language model to guide the expressive speech model in generating context-aligned speech. 
User studies show that V-CASS enhances emotional and attitudinal resonance, as well as user audio-visual understanding and engagement, with 74.68\% of participants preferring the system. 
Finally, we explore the potential of our method in helping blind and low-vision users navigate web videos, improving universal accessibility. 
\end{abstract}


\section{Introduction}
Video-centric social networks, such as TikTok, 
are becoming increasingly popular for featuring a wide range of short videos, 
including life vlogs, storytelling, and movie commentary.
AI technologies, such as automatic video commentary and captioning systems, can assist content creators in quickly producing video materials~\cite{apostolidis_video_2021, islam_exploring_2021, amirian_automatic_2020}.
These systems can generate real-time or post-processed captions or descriptions for uploaded videos, 
helping the audience to better understand the video content.
Video-accompanied captions and descriptions provide detailed information about the visual features, 
for example, what is happening, who is in the scene, and where it takes place.
In addition to the \textit{factual information} contained in visuals, 
videos also include a series of \textit{para-linguistic cues}, such as atmosphere, emotion, and attitude. 
These elements are conveyed through the visual context, encompassing aspects such as color tones, brightness, lines, and camera techniques~\cite{liu_paralinguistic_2023}.
To effectively convey the intended message of the video, 
it is crucial to jointly present both factual information and para-linguistic cues, 
as the combination of both factors practically contributes to the audience experience and promotes insightful content engagement.
However, current automatic video commentary and captioning systems may overlook the rich para-linguistic cues present in visuals, 
leading to an incomplete or distorted understanding.

\begin{figure*}[ht]
    \centering
    \includegraphics[width=\textwidth]{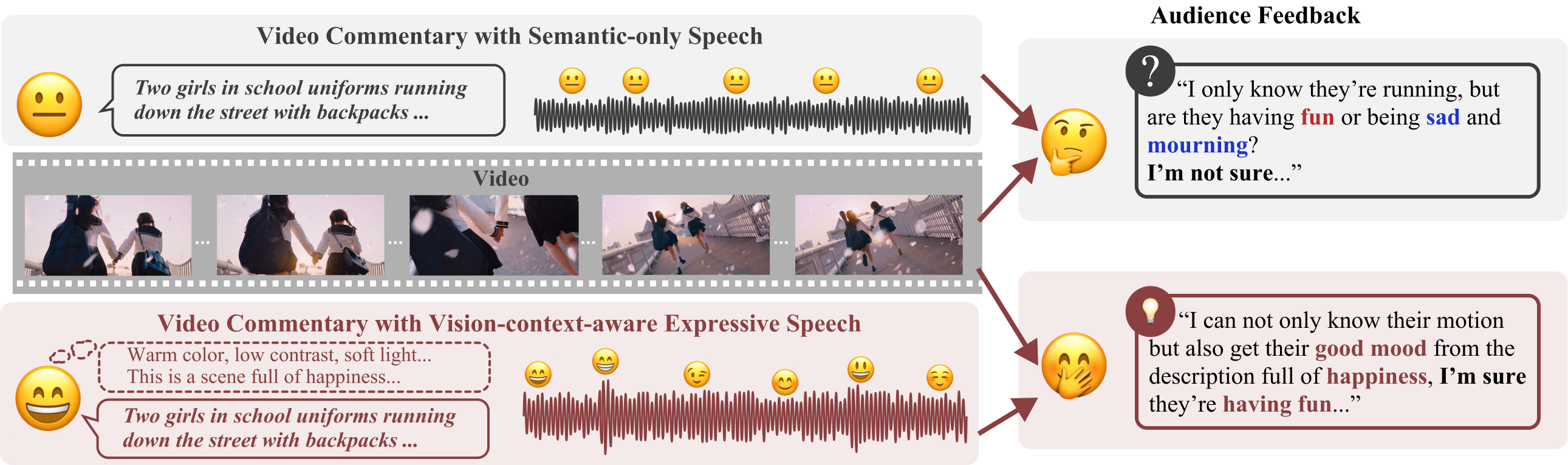}
    \caption{Potential effects on audience comparing semantic-only speech with vision-context-aware expressive speech in video commentary tasks.
Semantic-only speech lacks emotional and attitudinal information, which may lead to ambiguity and reduced engagement. 
In contrast, adding expressive speech that is aligned with visual para-linguistic cues can help enhance emotional resonance and improve the overall understanding of video content for audiences.}
\vspace{-5px}
    \label{fig:preview}
\end{figure*}

Expressive speech can be a direct way to convey para-linguistic cues. 
Consistent with human common sense, the varying aspects of vocal properties reflect distinct aspects of daily communication.
Tone and pitch reveal emotions such as joy or sadness, 
while intonation indicates whether a statement is a question or an assertion~\cite{scherer_vocal_2003}.
Rhythm and pace influence urgency, faster speech suggests immediacy, whereas slower delivery emphasizes important points. 
Lastly, volume signals confidence, 
louder speech implies assertiveness, while softer speech creates intimacy~\cite{chen_does_2022, steffman_rhythmic_2021}.
These speech features illustrate its ability to express complex para-linguistic cues embedded in videos. 
In fact, in video commentary, 
the alignment of visual and vocal elements significantly enhances the audience experience by providing not only semantic information but also a more holistic sensory experience~\cite{cacioppo_emotion_2007, lindquist_language_2021}.
\textit{Vision-context-aware expressive speech}, which synthesizes speech conditioned on visual cues, is crucial for enhancing audience engagement and understanding.
For example, Figure~\ref{fig:preview} illustrates the potential effects on audience comprehension when factual information expression in video commentary is aligned with two types of speech: 1) semantic-only speech, which conveys only the literal meaning without emotional expressive cues, and 2) vision-context-aware expressive speech.

Building on these insights, we explore the synthesis of vision-context-aware expressive speech. 
Firstly, we conduct a formatting study that demonstrates such speech significantly enhances user understanding and engagement, 
whereas misaligned speech distorts the interpretation of the video content. 
In these experiments, both vision-context-aware expressive speech and semantic-only speech are provided by professional voice actors.
Moreover, considering the importance of vision-context-aware expressive speech in video commentary, 
we propose a method called vision-context-aware speech synthesis (V-CASS).
Our method fuses multiple foundation models, 
including a vision-language model (VLM) to extract visual para-linguistic cues and a knowledge-infused large language model (LLM) to guide the expressive speech synthesis model in generating context-aligned speech.
User studies show that V-CASS enhances emotional and attitudinal resonance, as well as the users' audio-visual understanding and engagement, with 74.68\% of participants preferring the system over the semantic-only synthesized speech.
A case study with five blind and low-vision (BLV) participants indicates a preference for vision-context-aware speech synthesized by V-CASS as the audio descriptions (AD)\footnote{Audio description, which provides verbal descriptions of visual content for accessibility, is closely related to video commentary but serves a distinct role in making visual media accessible to blind and low-vision users.} ~\cite{gao_audio_2024, lopez_audio_2018, snyder_audio_2005}. 
This highlights its potential to significantly enhance accessibility for BLV users.

In summary, our contributions to the topic are threefold: 
\begin{itemize}[noitemsep, topsep=0pt, partopsep=0pt,leftmargin=*]
    \item We conduct a formatting study demonstrating that vision-context-aware expressive speech significantly enhances user understanding and engagement compared to misaligned or semantic-only speech. 
    \item We propose a novel method called V-CASS, which fuses multiple foundation models to generate expressive and context-aligned speech for video commentary.
    \item We demonstrate the potential of V-CASS to help BLV users more effectively access web-based multimedia content.\\
\end{itemize}


\vspace{-5px}
\section{Related Works}
\label{sect_relate_works}

\noindent\textbf{Visual para-linguistics.}
Visual context elements such as color, lighting, and composition reveal para-linguistic cues~\cite{song_ambient_2019}. 
For example, kobayashi~\cite{kobayasi_colorist_1998} conducted a study on color psychology and proposed a color-image coordinate system that categorizes colors based on people's emotional responses to them.
Colors were divided into four general categories: soft, hard, cool, and warm, each capable of invoking distinct emotions. 
Bright red is associated with gentleness, while dark black-brown evokes feelings of thickness and heaviness. Kobayashi~\cite{kobayasi_colorist_1998} also suggested that visual scenes carry valuable emotional and attitudinal information. Similarly, other visual aspects such as lighting and composition play a critical role in conveying emotions~\cite{bordwell_film_2013}. 
In contrast, low-key lighting, with its deep shadows and stark contrasts, is linked to suspense and mystery, typical of thrillers and horror genres. 

\vspace{10px}
\noindent\textbf{Expressive speech synthesis. }
Expressive speech synthesis produces lifelike voices by incorporating para-linguistic cues, prosodic information, and speaker characteristics.
Emotion is critical in speech expressiveness. 
Emotional TTS~\cite{wang_tacotron_2017, oord_wavenet_2016, li_styletts_2023, casanova_yourtts_2022} aim to generate speech that not only conveys semantic content but also captures emotional tone. 
These systems typically rely on manually assigned labels or emotional information extracted from the provided transcript, offering limited control over the synthesis of emotional speech.
Recently, language model-based speech synthesis foundation models, 
such as InstructTTS~\cite{yang_instructtts_2023}, VoxInstruct~\cite{zhou_voxinstruct_2024}, TextrolSpeech~\cite{ji_textrolspeech_2024}, and PromptTTS~\cite{leng_prompttts_2023}, 
have demonstrated improved performance by using natural language instructions to guide speech synthesis, 
enabling more flexible, comprehensive, and multidimensional fine-grained control of speech style.
For example, VoxInstruct controls prosody, emotion, and scenarios.

\vspace{10px}
\noindent\textbf{Linking visual and vocal information.}
Vision-based TTS systems can effectively link visual and vocal information. 
These systems can generate speech by taking visual inputs into account, making it possible to create more expressive and context-aware speech outputs. 
For example, Yang et al.\cite{yang_what_2023} proposed a model that generates speech from 3D face shapes, while multimodal integration enhances emotion expression by incorporating visual and auditory cues~\cite{baltrusaitis_multimodal_2019,tsai_multimodal_2019}. 
MM-TTS can utilize emotional cues from facial images to control the style of the generated speech~\cite{guan_mm-tts_2024}.
However, much of this research focuses on the expression on one's face, neglecting the vision para-linguistic cues~\cite{goto_face2speech_2020,guan_mm-tts_2024,lee_imaginary_2023,lu_face-based_2021,yang_what_2023}. 
Incorporating vision context information allows the model to better interpret and understand nuanced emotions in speech, 
thereby substantially improving the realism and emotional accuracy of synthesized speech~\cite{li_eald-mllm_2024,hertz_prompt--prompt_2022}.

\vspace{5px}
\noindent\textbf{Vision and language foundation models}.
The foundation models, including VLM and LLM, have transformed the landscape of multimodal processing~\cite{bommasani_opportunities_2021}.
VLMs, such as CLIP~\cite{radford_learning_2021}and ViLT~\cite{kim_vilt_2021}, are capable of extracting both the semantic details (objects, actions, etc.) and the broader visual context (scene setup, relationships between objects) in images and videos~\cite{ranzato_video_2014}. 
This is crucial for video commentary systems because understanding the visual context allows for a more nuanced interpretation, beyond mere factual description.
LLMs have achieved significant advancements, demonstrating capabilities such as detailed contextual reasoning and handling complex queries across various domains~\cite{bommasani_opportunities_2021}. 
Incorporating expert knowledge into LLMs is crucial to improve their accuracy and reliability in domain-specific tasks, 
where general knowledge may not be sufficient. 
Chain-of-Thought (CoT)~\cite{zhang_escot_2024, wei_chain--thought_2022} is a popular and straightforward method for injecting expert knowledge into LLMs by breaking down complex tasks into intermediate reasoning steps, 
allowing the model to process and respond more effectively.

\vspace{5px}
\noindent\textbf{Appending sound to video}.
Advancements in multimodal systems have explored generating audio for silent videos to enhance the viewer experience. 
For instance, Ghose and Prevost~\cite{ghose_autofoley_2021} introduced \textit{AutoFoley}, a deep learning tool that synthesizes synchronized soundtracks for silent videos by recognizing actions and temporal relationships within video clips. 
Ruan et al.\cite{ruan_mm-diffusion_2022} proposed MM-Diffusion, a joint audio-video generation framework utilizing a sequential multi-modal U-Net to produce realistic videos with synchronized audio. 
Michelsanti et al.\cite{michelsanti_vocoder-based_2020} presented a method to synthesize speech from silent video of a talker using deep learning, mapping raw video frames to acoustic features and reconstructing speech with a vocoder synthesis algorithm. 
Furthermore, Xie et al.\cite{xie_sonicvisionlm_2024} introduced SonicVisionLM, a framework that generates a wide range of sound effects by leveraging vision-language models. 
Instead of generating audio directly from video, this approach uses the capabilities of powerful vision-language models to identify events within the video and suggest possible sounds that match the visual content.

\section{Formatting Study of Audiovisual Emotion Mapping and Effect}
\label{sect_experiments}

As discussed previously, the emotion of speeches can influence how we interpret videos, 
and expressive speech can convey emotional nuances that enhance the users' understanding.
However, prior studies have not fully explored the effects of alignments or mismatches between visual para-linguistic and speech emotion. 
To tackle this challenge, an expert interview and a comparative experiment are necessary to examine
the effects that different emotional expressions in speech may have on users' understanding of visuals in video commentary tasks.

\begin{figure*}[ht]
    \centering
    \includegraphics[width=1\linewidth]{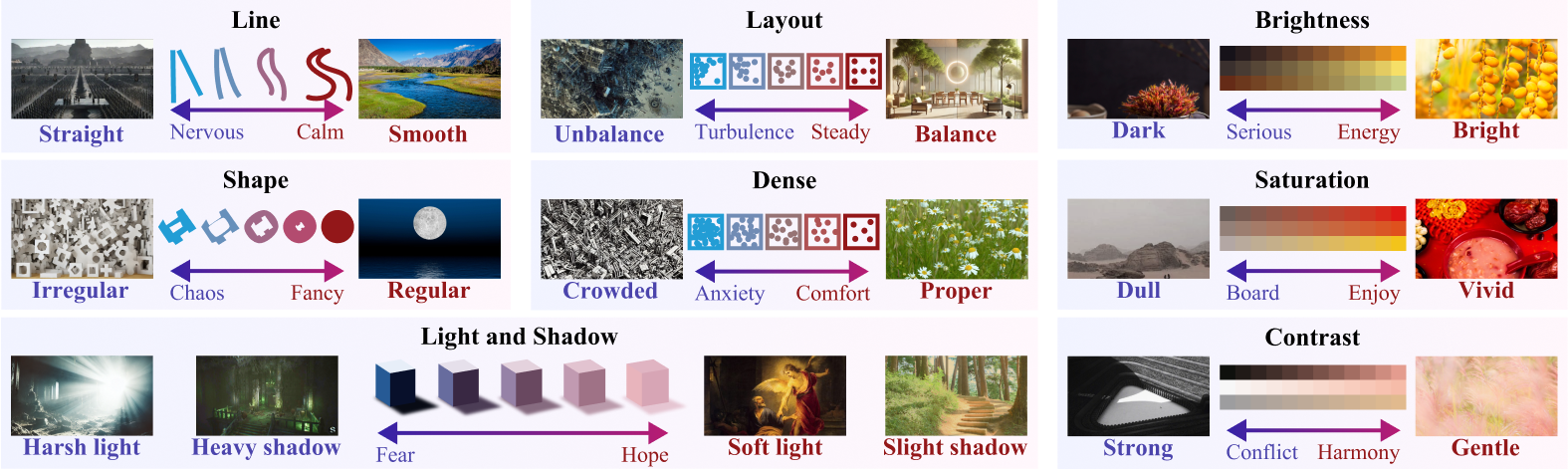}
    \vspace{-10px}
    \caption{This diagram illustrates samples of the vision-vocal para-linguistic mappings. The results are based on the problem-centered semi-structured expert interview with a professional voice actor, who indicated that visual para-linguistic information can be used to guide the vision-context-aware expression speech synthesis in video commentary tasks. For example, smooth lines can make people feel calm and peaceful, while straight lines may cause some nervousness. 
    Soft light and slight shadow may indicate the sign of hope, while harsh light and heavy shadow usually convey fearful feelings.
    }
    \vspace{-10px}
    \label{fig:visual-emotions}
\end{figure*}

\vspace{8px}
\noindent\textbf{Experts interview. }
We conducted an expert interview based on problem-centered~\cite{doringer_problem-centred_2021} and semi-structured methods with three professional voice actors who are experts in processing visual vocal affective mapping. 
The semi-structured interview, not only includes fixed questions but also allows for flexibility in adjusting to respondents' answers, was combined to gain deeper perspectives and insights from the experts (The interview materials, including the semi-structured interview protocol, are available at {\color{blue}\url{https://v-cass.github.io/V-CASS/}}). 
As Figure~\ref{fig:visual-emotions} shows, the results of the interview are organized into \textit{visual-to-vocal para-linguistic mapping knowledge}. 
It maps visual para-linguistic cues, e.g., color, lighting, layout, and composition, to corresponding emotional expressions, such as tone and pitch. 
The visual-to-vocal para-linguistic mapping knowledge is later applied to infuse the LLM with expert knowledge for processing visual vocal para-linguistic cues translation in section~\ref{sect_method}.\\

\noindent\textbf{Comparative Experiment settings}. 
To evaluate how speech emotion influences video commentary understanding, we prepared videos paired with speech recordings varying in emotional expression (Samples are available at {\color{blue}\url{https://v-cass.github.io/V-CASS/}}).
Participants evaluated the emotional tendency of the visual-vocal pairs.
Some visual cues such as color, composition, and lighting, are chosen for their potential to convey specific emotions and attitudes, either positive or negative. 
A subtle visual style ensures a balanced evaluation of visual and speech cues, 
enables participants to infer emotions from both visual and speech cues, and allows for a more focused evaluation of the emotional tone. 
Each video was paired with three speech settings: neutral (factual, emotionless), vision-context-aligned (emotionally consistent with the video), and emotionally contradictory (opposite emotion).
To ensure precise emotional delivery, either aligning with or contrasting against the video's visual tone, 
both the transcripts and audio recordings are produced by professional voice actors with extensive experience in expressing emotional nuances through speech. 
The transcripts are designed to be neutral and objective, focusing exclusively on the factual visual context to minimize interference from the semantics.
We introduce the pleasure scale from the \textit{pleasure-arousal-dominant} (PAD) model~\cite{mehrabian_pleasure-arousal-dominance_1996}, 
which measures the positivity or negativity of the emotional experience.
Research has discovered the direct connection of pleasure scale with affective responses~\cite{bakker_pleasure_2014}.
To simplify the task and avoid possible confusion, we convert it into a categorical rating system for participants.

A group of 30 participants are selected, primarily aged 18–25 (89\%), with 14 males and 15 females. 
Most participants were students (76\%), with some professionals from relevant fields.
Participants were chosen from three disciplines: arts and design (40\%), science (36\%), and human-computer interaction (23\%). 
Arts and design participants often engage with abstract emotional expressions, while those in science have less exposure to such content. 
Comparing the results from these two groups allowed us to ensure the consistency of emotional speech’s impact on visual understanding. 
Human-computer interaction participants, with their combined focus on user experience and human factors, added a complementary perspective.
This selection ensured diverse viewpoints and robust results.
Every video has an original video intent label as positive or negative.
Paired with speech, the overall visual-vocal tendency of video is evaluated as positive or negative as well.
Each video sample was rated by at least ten participants as a cross-labeler validation to ensure that the results are based on diverse perspectives, reducing the potential for individual bias.
We did not analyze specific tendencies in participants’ emotional misjudgments, because the primary objective of our study was to investigate whether the emotional quality of speech influences participants’ understanding. 
Analyzing specific misjudgment types (e.g., mistaking anger for sadness) was beyond the scope of this study. 
To ensure similar interpretations of emotional tendencies, participants also received standardized examples of different tendencies.


Experiment results verify the effectiveness of expressive speech in shaping the emotional tendency of videos.
As illustrated in Table~\ref{tab:pretest_emotion_tendency},
for each setting of speech, 
we denote the portion of positive/negative visual-vocal emotion tendency in videos with positive intent as PPT/PNT,
while the portion of positive/negative visual-vocal emotion tendency in videos with negative intent as NPT/NNT.
Since the videos are distributed balanced in the original intent (PI=NI), 
we calculate the portion of consistent/inconsistent samples to clearly demonstrate the effects of speech,
by focusing on the consistency between the original intent and the visual-vocal tendency.

\vspace{10px}
\noindent\textbf{Results and findings 1: Effects of expressive speech.}
When presented with neutral speech, which provided only semantic, emotionless contexts, 
participants' ability to correctly understand the emotional and contextual aspects of the video was relatively low, with a correct understanding rate of 65\%. 
The absence of emotional cues left them rely on the semantic content of the speech, making it difficult to infer missing emotional information from the video, thereby hindering their deeper comprehension. 
This underscores the limitations of semantic-based speech in the video commentary task, 
as the lack of emotional guidance may result in an incomplete and inconsistent understanding of the visual content.
Despite the original videos with no speech showing a 50-50 split of positive and negative intent,
when given a video with neutral speech, the distribution of positive and negative tendencies evaluated by human participants is not balanced.
Although the sample size was relatively even across the different intent, there might be some subjective variation in the analysis of visual para-linguistic cues between participants, which may have led to unbalanced test results.
In comparison, the presence of emotional cues in the speech (whether consistent or inconsistent) had a marked impact on human evaluation, as shown in the gray cell of Table~\ref{tab:pretest_emotion_tendency}.
Compared to neutral speech, the consistent value corresponding to visual-context-aligned speech is significantly higher, 79.95\% over 64.00\%, indicating the emotional context enhancement of the aligned speech.
While the inconsistent value corresponding to emotionally contradictory speech is significantly higher than that of neutral speech, 
indicating that the misaligned speech could cause a misunderstanding of the emotional context in the videos.

To conclude, in emotionally charged speech conditions, the emotional tone of the video is dominated by the emotional expression of the speech.  
For example, when exposed to speech with a positive tone, participants were more likely to judge the video as conveying a positive tone. 
On the contrary, when presented with a negatively charged speech, participants tended to interpret the video in a more negative light. 
Emotional cues in speech act as valuable complements to the semantic content, bridging gaps left by purely semantic descriptions. 
Appropriate emotional expression in speech can help participants interpret the visual material more clearly and enable them to make more confident judgments.\\

\begin{table*}[htbp]
\centering
\caption{Participants evaluated the visual-vocal emotion tendency for each video-speech pair. The first P/N represents the original visual intent (positive/negative), while the second P/N indicates the tendency of the evaluation results. 
PPT refers to videos with positive intent and positive tendency (consistent), PNT indicates positive intent but negative tendency (inconsistent), NPT represents negative intent but positive tendency (inconsistent), and NNT reflects negative intent with negative tendency (consistent). 
The consistency rate is calculated as (\text{PPT} + \text{NNT}) / (\text{PI} + \text{NI}), while the inconsistency rate is (\text{PNT} + \text{NPT}) / (\text{PI} + \text{NI}). 
For example, when paired with neutral speech, 70.60\% of videos with positive intent exhibit positive tendency, while 29.40\% show negative tendency. 
Combining both intents, the overall consistency rate is (70.60\% + 57.40\%) / 2 = 64.00\%.
}
\resizebox{\linewidth}{!}{
\begin{tabular}{ccccccc}
\toprule

& \multicolumn{2}{c}{\textbf{Positive Intent (PI)}} & \multicolumn{2}{c}{\textbf{Negative Intent (NI)}} & \multirow{2}{*}{\textbf{Consistent}} & \multirow{2}{*}{\textbf{Inconsistent}}\\
& \textbf{Positive (PPT)} & \textbf{Negative (PNT)} & \textbf{Positive (NPT)} & \textbf{Negative (NNT)} \\
\midrule

\multirow{2}{*}{\textbf{Neutral speech}} & \multirow{2}{*}{70.60\%} & \multirow{2}{*}{29.40\%} & \multirow{2}{*}{42.60\%} & \multirow{2}{*}{57.40\%}  & \multirow{2}{*}{64.00\%} & \multirow{2}{*}{36.00\%}  \\
& & & & & & \\
\hdashline
{\textbf{Vision-context-aligned}}& & & & &\cellcolor{gray!25} & \\
{\textbf{speech}} & \multirow{-2}{*}{\textbf{ 91.10\%}} & \multirow{-2}{*}{ 9.90\%} & \multirow{-2}{*}{ 31.20\%} & \multirow{-2}{*}{\textbf{ 68.80\%}} & \multirow{-2}{*}{\cellcolor{gray!25} \textbf{79.95\%}} & \multirow{-2}{*}{20.05\%}  \\
\hdashline
{\textbf{Emotionally contradictory }}& & & & & & \cellcolor{gray!25}\\
{\textbf{speech}} & \multirow{-2}{*}{64.90\%} & \multirow{-2}{*}{\textbf{35.10\%}} & \multirow{-2}{*}{\textbf{48.90\%}} & \multirow{-2}{*}{51.10\%}  & \multirow{-2}{*}{58.00\%} &\multirow{-2}{*}{\cellcolor{gray!25}  \textbf{42.00\%}} \\
\bottomrule
\end{tabular}
}
\label{tab:pretest_emotion_tendency}
\end{table*}

\noindent\textbf{Results and findings 2: Importance of emotion alignment. }  
While emotional cues do enhance understanding, the alignment between speech and visual content is critical. 
The comparison between consistent and inconsistent speech conditions highlights the importance of this alignment. 
In contrast, the inconsistent speech condition, where the emotional tone of the description conflicted with the video’s content, led to a significantly lower correct understanding rate of 60\%. 
This demonstrates that emotional alignment in the speech not only enhances comprehension but also helps participants interpret visual content more accurately.
The conflicting emotional information disrupted participants' ability to interpret the video correctly, causing them to form incorrect conclusions about its emotional tone and context.
In conclusion, the preliminary experiment shows that neutral speech, which lacked emotional cues, left participants with insufficient information to accurately interpret the videos, demonstrating the limitations of semantic-only speech. 
the critical role that vision-context-aware expressive speech plays in enhancing participants' understanding and engagement. 
In contrast, emotionally charged speech (both consistent and inconsistent) significantly influenced participants' comprehension. 
While consistent speech improved understanding by filling in gaps left by the semantic information, inconsistent speech actively misled participants, causing misunderstandings of the video's content.\\


\vspace{-10px}
\section{Generating Vision-context-aware Speech}
\label{sect_method}
Previous sections have concluded that vision-context-aligned expressive speech can help users better understand information from visuals. 
In this section, we introduce the V-CASS method, designed to generate speech that is semantically and emotionally aligned with visual context.\\


\begin{figure}[t]
    \centering
    \includegraphics[width=0.485\textwidth]{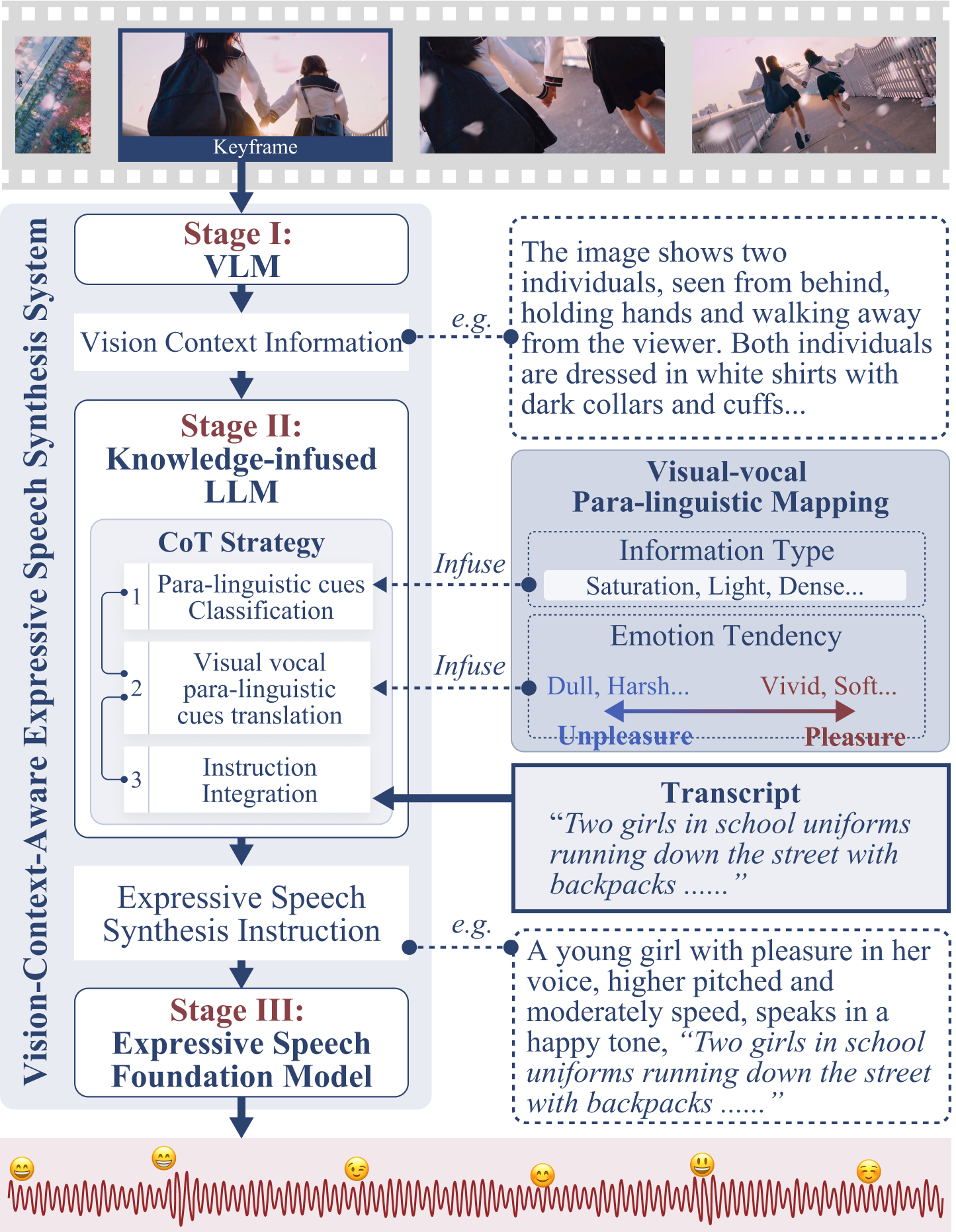}
    \caption{Overview of V-CASS method for synthesizing vision-context-aware expressive speech. V-CASS consists of three stages. 
Stage I extracts visual para-linguistic cues using a VLM, 
capturing key visual attributes. In Stage II, a knowledge-infused LLM translates these cues into vocal expressions. 
Stage III synthesizes context-aware expressive speech using the expressive speech foundation model.
}
\vspace{-10px}
    \label{fig:pipeline}
\end{figure}



\noindent\textbf{Stage I: Vision para-linguistic cues analysis. }
The first stage in the V-CASS method is the analysis of the vision para-linguistic cues. 
As discussed in Section~\ref{sect_relate_works}, VLMs are highly efficient and comprehensive in analyzing various aspects of the visual content. 
We employ the state-of-the-art VLM $f_V$, Google’s Gemini~\cite{gemini_team_gemini_2023}, instructed by VLM prompt $\mathcal{P}_v$ to extract the most comprehensive information $\mathcal{C}_v$ possible from keyframes $\mathcal{I}_{\text{input}}$ of the video $\mathcal{V} = \{\mathcal{I}_1, \cdots, \mathcal{I}_n\}$ for $n \in \mathbb{N}$. 
The output text describes visual attributes such as visual effects (e.g., bokeh, lighting), color composition, environmental context (e.g., crowded outdoor space, background elements), 
and the emotional tone conveyed by the scene (e.g., warmth, festivity, or energy).
This process can be formulated as:
\begin{equation}
    \mathcal{C}_v = f_V(\mathcal{I}_{\text{input}},\mathcal{P}_v).
\end{equation}

\noindent\textbf{Stage II: Translating from visual to vocal para-linguistic cues. }
Based on the visual attributes described in Stage I, 
Stage II applies a knowledge-infused LLM to translate visual para-linguistic cues $\mathcal{C}_v$ into vocal para-linguistic expressions. 
We utilize GPT-4o\footnote{\url{https://openai.com/index/hello-gpt-4o/}}, a state-of-the-art model, as the base LLM $f_L$, 
and infuse expert knowledge, specifically visual-to-vocal para-linguistic mapping knowledge 
as discussed in Section~\ref{sect_experiments}, into the LLM using a CoT strategy with the prompt $\mathcal{P}_l$. 
As Fig.~\ref{fig:visual-emotions} shows, the expert knowledge $\mathcal{K}_{\text{expert}}$, gathered through interviews with professional voice actors who have solid experience in processing the audiovisual emotions, 
and research in visual-emotional perception provides the foundation for interpreting how visual elements such as color, lighting, and other contextual cues translate into specific emotional tones.
CoT strategy allows the LLM to systematically break down complex visual data and identify the elements most relevant for emotional inference. 
In the first step, visual para-linguistic cues $\mathcal{C}_v$ are input into the system. 
Leveraging expert knowledge $\mathcal{K}_{\text{expert}}$, the LLM classifies these cues to extract relevant contextual information, which forms the foundation for subsequent visual affective analysis. 
Following this, the selected visual cues are mapped to their corresponding emotional states.
Finally, the emotional information and transcript $\mathcal{T}_{\text{tran}}$ are integrated and converted into structured instructions $\mathcal{T}_{\text{ins}}$, guiding the synthesis of context-aligned speech.
Finally, the LLM integrates the given transcript with the vocal expressiveness described above, 
forming an instruction containing both semantic and emotional information to guide the expressive speech synthesis in the next stage. 
Stage II can be expressed as:
\begin{equation}
    \mathcal{T}_{\text{ins}} = f_L(\mathcal{C}_v \cap \mathcal{K}_{\text{expert}}, \mathcal{T}_{\text{tran}}, \mathcal{P}_l).
\end{equation}

\vspace{10px}
\noindent\textbf{Stage III: vision-context-aware speech synthesis. }
This stage focuses on synthesizing vision-context-aware speech $\mathcal{A}_{\text{output}}$ based on the expressive speech synthesis instruction $\mathcal{T}_{\text{ins}}$, which is organized in Stage II. 
In this stage, we utilize VoxInstruct~\cite{zhou_voxinstruct_2024} as introduced in Section~\ref{sect_relate_works}, 
which can process speech synthesis instructions $\mathcal{T}_{\text{ins})}$ that contain both semantic information (transcript) $\mathcal{T}_{\text{tran}}$ and vocal para-linguistic cues $\mathcal{C}_v$. 
This stage can be represented by:
\begin{equation}
    \mathcal{A}_{\text{output}} = f_S (\mathcal{T}_{\text{ins}}),
\end{equation}
where $f_S$ represents a instruct-to-speech model and $\mathcal{A}_{\text{output}}$ denotes the output vision-context-aware speech. 
An example of such an instruction is shown in Figure~\ref{fig:pipeline}.
In summary, V-CASS consists of three stages. 
Stage I extracts visual para-linguistic cues using a VLM, 
capturing key visual attributes. In Stage II, a knowledge-infused LLM translates these cues into vocal expressions, 
and in Stage III,  synthesize context-aware expressive speech using VoxInstruct.

\section{User Studies}
\label{sect_user_studies}

We conduct two experiments designed to measure both the emotional inference capabilities of our method and its performance in generating expressive speech. 
These experiments aim to validate V-CASS's ability to extract emotional cues from visual content and demonstrate its impact on enhancing the user experience through appending vision-context-aware speech.
We invite the 30 participants from the formatting study to take part in the user study to ensure consistency and minimize variability caused by individual differences. 
Their prior experience with the experimental setup and aims could reduce the learning process, allowing the user study to focus on evaluating the V-CASS’s effectiveness, ensuring that observed results reflected the method’s impact rather than differences in participant characteristics, enhancing the study’s validity.\\

\begin{table*}[h!]
\centering
\caption{Textual similarity calculated by Sentence-BERT comparing the knowledge-infused LLM and original LLM with participants' results for each video. The emotional descriptions generated by the LLM with expert knowledge showed a significantly higher similarity to the human-provided emotional descriptions.}
\begin{tabular}{@{}cccccccccccc@{}}
\toprule
\textbf{\textit{Vedio ID}} & \textbf{\textit{01}} & \textbf{\textit{02}} & \textbf{\textit{03}} & \textbf{\textit{04}} & \textbf{\textit{05}} & \textbf{\textit{06}} & \textbf{\textit{07}} & \textbf{\textit{08}} & \textbf{\textit{09}} & \textbf{\textit{10}} & \multicolumn{1}{c}{ \cellcolor{gray!25}  \textbf{\textit{Average}}}\\
\midrule
{\textbf{Knowledge-infused LLM}} & {0.67} & {\textbf{0.75}} & {\textbf{0.73}} & {\textbf{0.65}} & {\textbf{0.61}} & {\textbf{0.70}} &  {\textbf{0.80}} & {0.67} & {\textbf{0.73}} & {\textbf{0.67}} &\multicolumn{1}{c}{\cellcolor{gray!25} \textbf{0.70}}\\
\hdashline
{\textbf{Non-knowledge-infused LLM}} & {\textbf{0.68}} & {0.64} & {0.65} & {0.61} & {0.52} & {0.69} & {0.57} & {\textbf{0.68}} & {0.61} &{0.59} &\multicolumn{1}{c}{\cellcolor{gray!25} 0.62}\\
\bottomrule
\end{tabular}
\label{tab:similarity-result}
\end{table*}
\noindent\textbf{Efficiency of knowledge-infused LLM. }
In the first experiment, we compare the emotional inference abilities of an LLM integrated with expert knowledge against an LLM without such expert input, as well as a group of human participants. 
The experiment involved a set of ten videos without audio, each presenting a distinct emotional tone. All videos used for experiments were screened according to the same standards as the videos used in the formatting study. A total of 30 participants were asked to write down the emotional information for videos showed to them. Each video sample was evaluated by at least 20 participants. Simultaneously, the same videos are analyzed by two LLMs: the knowledge-infused LLM, and the other without expert injection.
Textual descriptions from participants, the knowledge-infused LLM, and general LLM were compared using Sentence-BERT~\cite{reimers_sentence-bert_2019} to calculate similarity. 
As Table~\ref{tab:similarity-result} shows, the similarity score of knowledge-infused LLM remains higher than non-knowledge-infused LLM, demonstrating that the emotional descriptions generated by the LLM with expert knowledge showed a significantly higher similarity to the human-provided emotional descriptions compared to the LLM without expert knowledge. 
Although the same emotion description template was provided to the participants and two version of LLMs, the non-knowledge-infused LLM performance exhibited consistent differentiation. 
This highlights the advantage of incorporating expert knowledge into the LLM to better capture and align with human-provided emotional descriptions.
The “chain-of-thinking” reasoning in V-CASS enables sentiment-enhanced descriptions closer to human interpretations, outperforming the standard LLM.


\vspace{10px}
\noindent\textbf{Vision-context-aware expressive speech synthesis enhancing user engagement. }
The second experiment is designed to compare the participants' preferences between emotionally expressive speech generated by our V-CASS method and neutral speech generated by a basic TTS. 
A set of ten videos is selected, with two different speech outputs generated for each: one using the \textit{TTS Maker} online tool, which provided neutral speech without emotional para-linguistic cues, and the other using our vision-context-based speech synthesis method.
Each video, accompanied by one of the two speech versions, was shown to the same group of 30 participants. 
After watching each video, participants were asked to evaluate the clarity and expressiveness of the speech and to state which one they preferred. 
Participants were specifically asked which speech conveyed the emotional content of the video more effectively and which one they found more intriguing.

The experiment results indicate that 74.68\% of the participants preferred the expressive speech generated by V-CASS. 
They reported that the speech generated by the V-CASS method better captured the emotional nuances of the video, resulting in a more lucid and engaging viewing experience overall.
This pronounced preference for our method highlights its effectiveness in generating emotionally expressive and contextually aligned speech that enhances users' comprehension and engagement with vision para-linguistic.
In conclusion, participants preferred V-CASS-generated videos with emotionally expressive speech. 
Speech with emotion is more attuned to human cognitive characteristics and can better capture the nuances of emotions, which not only conveys richer information but also makes the video more appealing.

\vspace{5px}
\section{Case studies on BLVs}
\label{sect_case_studies}
To assess the potential of vision-context-aware speech synthesis, we conduct a subjective experiment with five blind and low-vision (BLV) participants.
Each of the BLV participants was presented with both neutral speech and vision-context-aligned speech, and a silent video. 
They were requested to consider the speech as audio descriptions (AD) for the video, with the assumption that they were viewing an accessible video with the assistance of the AD. 
Participants were then required to respond to questions related to the video's narrative style, potential emotional tendencies, and other related aspects based on their understanding.
After the test, a post-experiment survey gathered participants' subjective feedback. 

The experiment results demonstrated that participants were more accurate in inferring the video context when presented with context-aligned expressive speech, compared to neutral speech. 
The post-experiment feedback further revealed that BLV participants preferred context-aligned expressive speech, for it could provide an engaging and more immersive experience. 
To conclude, similar to human-generated AD, context-aligned expressive speech with more emotional expressiveness improved the participants' comprehension of the visual context and elicited a stronger emotional response.
These findings highlight the broader applicability of V-CASS, suggesting that vision-context-align expressive speech can significantly enhance user experiences for accessible applications.

\section{Discussion}
\label{sect_discussion}
Compared with ordinary TTS systems, V-CASS does not simply improve speech synthesis technology but proposes a novel method of synthesizing speech. Traditional TTS models rely on human instruction, which is inefficient and difficult to ensure consistency. At the same time, TTS rarely considers visual content, and even with image control, it focuses more on isolated elements such as facial expressions and ignores the broader visual context.
V-CASS automatically generates speech with high stability and quality. It aligns speech with the emotional tone of the visual scene and is not limited to facial expressions. This effectively improves emotional consistency while ensuring semantic accuracy.

\section{Conclusion}
\label{sect_conclusion}
In this paper, we introduced V-CASS, 
a method that leverages visual context to guide expressive speech synthesis. 
It enhances user comprehension and engagement by aligning speech with the para-linguistic cues present in videos. 
Our formatting stduy demonstrate that vision-context-aware expressive speech significantly improves understanding compared to semantic-only speech. 
User studies also highlight its potential to enhance accessibility for BLV users by providing richer, 
more vivid AD that conveys the nuances of visuals. 
By offering more immersive and expressive speech, 
V-CASS shows strong potential for boosting content creation in video-centric social networks, 
making multimedia experiences more inclusive and engaging for the audience.


\bibliographystyle{IEEEtran}
\bibliography{main}
\end{document}